\documentstyle[emulateapj]{article}

\submitted{to appear in the Astrophysical Journal, Letters, vol. 528}

\lefthead{Hachisu et al.}
\righthead{Light Curve of U Scorpii Outburst}

\begin{document}

\title{A THEORETICAL LIGHT-CURVE MODEL FOR THE 1999 OUTBURST OF 
U SCORPII}

\author{Izumi Hachisu}
\affil{Department of Earth Science and Astronomy, 
College of Arts and Sciences, University of Tokyo,
Komaba, Meguro-ku, Tokyo 153-8902, Japan; hachisu@chianti.c.u-tokyo.ac.jp}

\author{Mariko Kato}
\affil{Department of Astronomy, Keio University, 
Hiyoshi, Kouhoku-ku, Yokohama 223-8521, Japan; mariko@educ.cc.keio.ac.jp}
\and 

\author{Taichi Kato and Katsura Matsumoto}
\affil{Department of Astronomy, Kyoto University, 
Kitashirakawa, Sakyo-ku, Kyoto 606-8502, Japan; 
tkato@kusastro.kyoto-u.ac.jp, katsura@kusastro.kyoto-u.ac.jp}

% Notice that each of these authors has alternate affiliations, which
% are identified by the \altaffilmark after each name.  The actual alternate
% affiliation information is typeset in footnotes at the bottom of the
% first page, and the text itself is specified in \altaffiltext commands.
% There is a separate \altaffiltext for each alternate affiliation
% indicated above.

%\altaffiltext{1}{Visiting Astronomer, Cerro Tololo Inter-American 
%Observatory.  
%CTIO is operated by AURA, Inc.\ under contract to the National Science
%Foundation.} 

% The abstract environment prints out the receipt and acceptance dates
% if they are relevant for the journal style.  For the aasms style, they
% will print out as horizontal rules for the editorial staff to type
% on, so long as the author does not include \received and \accepted
% commands.  This should not be done, since \received and \accepted dates
% are not known to the author.

\begin{abstract}
A theoretical light curve for the 1999 outburst of U Scorpii is 
presented in order to obtain various physical parameters 
of the recurrent nova. 
Our U Sco model consists of a very massive white dwarf (WD) 
with an accretion disk and a lobe-filling, slightly evolved, 
main-sequence star (MS).  The model includes a reflection effect 
by the companion and the accretion disk together with 
a shadowing effect on the companion by the accretion disk.  
The early visual light curve (with a linear phase of $t \sim 1-15$ days 
after maximum) is well reproduced by a thermonuclear runaway model 
on a very massive WD close to the Chandrasekhar limit
($M_{\rm WD}= 1.37 \pm 0.01 ~M_\odot$), in which optically thick 
winds blowing from the WD play a key role in determining the nova duration.
The ensuing plateau phase ($t \sim 15-30$ days) is also reproduced 
by the combination of a slightly irradiated MS 
and a fully irradiated flaring-up disk with a radius 
$\sim 1.4$ times the Roche lobe size.
The cooling phase ($t \sim 30-40$ days) is consistent with 
a low hydrogen content $X \approx 0.05$ of the envelope 
for the $1.37 M_\odot$ WD.  The best fit parameters are the WD mass 
$M_{\rm WD} \sim 1.37 ~M_\odot$, the companion mass 
$M_{\rm MS} \sim 1.5 M_\odot$ ($0.8-2.0 M_\odot$ is acceptable),
the inclination angle of the orbit $i \sim 80 \arcdeg$, and 
the flaring-up edge, the vertical height of which is $\sim 0.30$ 
times the accretion disk radius.
The duration of the strong wind phase ($t \sim 0-17$ days)
is very consistent with the BeppoSAX supersoft X-ray detection
at $t \sim 19-20$ days because supersoft X-rays are self-absorbed 
by the massive wind.  The envelope mass at the peak is estimated to be 
$\sim 3 \times 10^{-6} M_\odot$, which is indicating 
an average mass accretion rate 
$\sim 2.5 \times 10^{-7} M_\odot$ yr$^{-1}$ 
during the quiescent phase between 1987 and 1999. 
These quantities are exactly the same as those predicted 
in a new progenitor model of Type Ia supernovae. 
\end{abstract}

% The different journals have different requirements for keywords.  The
% keywords.apj file, found on aas.org in the pubs/aastex-misc directory, 
% contains a list of keywords used with the ApJ and Letters.  These are 
% usually assigned by the editor, but authors may include them in their 
% manuscripts if they wish. 


% That's it for the front matter.  On to the main body of the paper.
% We'll only put in tutorial remarks at the beginning of each section
% so you can see entire sections together.

% In the first two sections, you should notice the use of the LaTeX \cite
% command to identify citations.  The citations are tied to the
% reference list via symbolic KEYs.  We have chosen the first three
% characters of the first author's name plus the last two numeral of the
% year of publication.  The corresponding reference has a \bibitem
% command in the reference list below.
%
% Please see the AASTeX manual for a more complete discussion on how to make
% \cite-\bibitem work for you.   

\section{INTRODUCTION}
      U Scorpii is one of the well observed recurrent 
novae, characterized by the shortest recurrence period 
($\sim 8$ yr), the fastest decline of its light curve
(0.6 mag day$^{-1}$), and its extremely helium-rich ejecta 
(He/H$\sim 2$ by number; see, e.g., \cite{web87} for summary).
Historically, the outbursts of U Sco were observed 
in 1863, 1906, 1936, 1979, 1987, and the latest in 1999. 
Especially, the 1999 outburst was well observed 
from the rising phase to the cooling phase 
by many observers (e.g., \cite{mun99}), including eclipses 
(\cite{mat99}), thus providing us with a unique opportunity to construct 
a comprehensive model of U Sco during the outburst.  
\par
     Our purpose in this Letter is to construct 
a detailed theoretical model for U Sco 
based on our light-curve analysis.  A part of
the method has been described in
Hachisu \& Kato (1999), in which they explain the second peak 
of T CrB outbursts, but we again briefly summarize it in \S 2, and
will fully describe it in a separate paper.
In \S 3, by directly fitting our theoretical light curve 
to the observational data,  
we derive various physical parameters of U Sco.
%, i.e.,
%the white dwarf (WD) mass,  ($M_{\rm WD}$), 
%the envelope mass ($\Delta M$) on the WD, 
%the hydrogen content ($X$) of the envelope,
%and various physical parameters of the binary,
%e.g., the size and thickness of the accretion disk. 
Discussions follow in \S 4, especially in relation to 
the recently proposed progenitor model of Type Ia
supernovae (SNe Ia).
%\par
%     The main results of our analysis are:\\
%1) The early visual light curve of a linear decay phase 
%($t \sim 1-10$ d) is well reproduced 
%by a thermonuclear runaway model on a $1.36 M_\odot$ WD 
%with an envelope mass of $\sim 3 \times 10^{-6} M_\odot$ 
%at the peak, which is indicating a mass transfer rate 
%of $\dot M_{\rm acc} \sim 2-3 \times 10^{-7} M_\odot$ yr$^{-1}$.\\
%2) The ensuing plateau phase ($t \sim 15-30$ d) is also 
%reproduced by the combination of the slightly irradiated 
%MS star and the irradiated flaring-up disk with a large radius 
%of $\sim 1.2$ times the Roche lobe size.\\
%3) The inclination angle of the orbit is $i \sim 76 \arcdeg$, and 
%the flaring-up edge of $\sim 0.35$ times the accretion disk radius.
%4) The final decay phase ($t \sim 30-40$ d) is consistent with 
%a low hydrogen content of $X \approx 0.1$ for the $1.36 M_\odot$ 
%WD because of a very fast extinguishment of hydrogen shell burning.\\

\section{THEORETICAL LIGHT CURVES}
     Schaefer (1990) and Schaefer \& Ringwald (1995) observed 
eclipses of U Sco in the quiescent phase and determined 
the orbital period ($P= 1.23056$ days) and 
the Ephemeris (HJD 2,451,235.777 $+$ 1.23056$E$) at the
epoch of mideclipse.  It is very likely 
that the companion is a slightly evolved main-sequence star (MS)
that expands to fill its Roche lobe after a large
part of the central hydrogen is consumed.
The phase duration ($\Delta \varphi \sim 0.1$) of the primary eclipses 
in quiescent phase indicates an inclination angle of $i \sim 80\arcdeg$ 
(e.g., \cite{war95}). 
\par
     Our model is shown graphically in Figure \ref{usco_fig80_w30}.  
We have revised Kato's (1990) U Sco light-curve models 
in the following two ways: (1) the opacity has been changed
from the Los Alamos opacity (Cox, King, \& Tabor 1973) 
to the OPAL opacity (\cite{igl96}) and
(2) the reflection/irradiation effects both of the companion star and
the accretion disk are introduced 
in order to follow the light curve during
the entire phase of the outburst.  The visual light curve is 
calculated from three components of the system: 
(1) the white dwarf (WD) photosphere, 
(2) the MS photosphere, and
(3) the accretion disk surface.
\placefigure{usco_fig80_w30}

\subsection{Decay Phase of Novae}
     In the thermonuclear runaway model
(e.g., Starrfield, Sparks, \& Shaviv 1988), WD envelopes 
quickly expand to $\sim 10-100 ~R_\odot$ or more and 
then the photospheric radius gradually shrinks to the original 
size of the white dwarfs.
% (e.g., $\sim 0.0032 ~R_\odot$ for 
%$M_{\rm WD}= 1.37 ~M_\odot$). 
Correspondingly, the optical luminosity reaches its maximum 
at the maximum expansion of the photosphere and then 
decays toward the level in the quiescent phase, 
keeping the bolometric luminosity almost constant.
Since the WD envelope reaches a steady-state after maximum,
we are able to follow the development by a unique sequence 
of steady-state solutions %($\Delta M$) 
as shown by Kato \& Hachisu (1994). 
Optically thick winds, blowing from the WD in the decay
phase of novae, play a key role in determining the nova duration
because a large part of the envelope mass is quickly blown off
in the wind.
%and occur when
%the WD photosphere is larger than $\sim 0.05-0.1 R_\odot$, i.e.,
%when the photospheric temperature is lower than
%$\log T_{\rm ph} \lesssim 5.5$, because the wind is driven by 
%a strong peak at $\log T \sim 5.2$ in OPAL opacity (e.g., \cite{igl96}).  
%\par
%     Assuming that the solar abundance of heavy elements
%($Z=0.02$), we have calculated such sequences 
%for WDs with various masses of $M_{\rm WD}= 1.3$, 
%1.35, 1.36, 1.37, and $1.377 M_\odot$, and obtained the optical 
%light curves for the decay phase of TNR events.  
%The numerical methods and various assumptions are 
%similar to those in Hachisu \& Kato (1999).
\par
     In the decay phase, the envelope structure at any given time
is specified by a unique solution.
The envelope mass $\Delta M$
is decreasing because of both the wind mass loss 
at a rate of $\dot M_{\rm wind}$ 
and the hydrogen shell burning 
at a rate of $\dot M_{\rm nuc}$, i.e., 
\begin{equation}
{{d} \over {d t}} \Delta M = \dot M_{\rm acc} - 
\dot M_{\rm wind} - \dot M_{\rm nuc},
\label{dmdt_envelope_mass}
\end{equation}
where $\dot M_{\rm acc}$ is the mass accretion rate of the WD.
By integrating equation (\ref{dmdt_envelope_mass}), 
we follow the development of the envelope 
and then obtain the evolutionary changes of physical quantities such as 
the photospheric temperature $T_{\rm ph}$, the photospheric radius
$R_{\rm ph}$, the photospheric velocity $v_{\rm ph}$, the wind mass loss 
rate $\dot M_{\rm wind}$, and the nuclear burning rate
$\dot M_{\rm nuc}$.
%as a unique function of the envelope mass $\Delta M$.
When the envelope mass decreases to below the critical mass,
the wind stops, and after that, the envelope mass is decreased 
only by nuclear burning.  
When the envelope mass decreases further,
hydrogen shell-burning disappears,
%above breaks down at the minimum envelope mass of the sequence.
%, i.e., 
%$\Delta M_{\rm min}= 1.07 \times 10^{-7} M_\odot$ 
%for the sequence of $M_{\rm WD}=1.37 M_\odot$, $X=0.05$, and $Z=0.02$.  
and the WD enters a cooling phase.

\subsection{White Dwarf Photosphere}
     After the optical peak, the photosphere shrinks 
with the decreasing envelope mass mainly because of the wind mass loss. 
Then the photospheric temperature increases and the visual light 
decreases because the main emitting region moves blueward (to UV 
then to soft X-ray).  
We have assumed a blackbody photosphere of the white 
dwarf envelope and estimated visual magnitude 
of the WD photosphere with a window function given by 
Allen (1973).   The photospheric surface is divided into 
32 pieces in the latitudinal angle %($\Delta \theta = \pi/32$) 
and into 64 pieces in the longitudinal angle %($\Delta \phi = 2 \pi/64$) 
as shown in Figure \ref{usco_fig80_w30}. 
%Then, the contribution of each piece is summed up.
%by considering the inclination angle to Earth.
For simplicity, we do not consider the limb-darkening effect.

\subsection{Companion's Irradiated Photosphere}
     To construct a light curve, we have also included 
the contribution of the companion star
irradiated by the WD photosphere.  
The companion star is assumed to be synchronously rotating 
on a circular orbit and 
its surface is also assumed to fill the inner 
critical Roche lobe, as shown in Figure \ref{usco_fig80_w30}.
Dividing the latitudinal angle into 
32 pieces %($\Delta \theta = \pi/32$) 
and the longitudinal angle into
64 pieces, %($\Delta \phi = 2 \pi/64$), 
we have also summed up
the contribution of each patch, but, for simplicity,
%by considering the inclination angle to Earth, 
we neglect both the limb-darkening effect and
the gravity-darkening effect of the companion star.
Here we assume that 50\% of the absorbed energy 
is reemitted from the companion surface with a blackbody 
spectrum at a local temperature.
The original (nonirradiated) photospheric temperature 
of the companion star is assumed to be $T_{\rm ph, MS} = 4600$ K
because of $B-V=1.0$ in eclipse (\cite{sch95}; see also \cite{joh92}).  
Two other cases of $T_{\rm ph, MS} = 5000$ and 4400 K have been 
examined but do not show any essential differences in the light curves. 
%If the accretion disk around the WD
%blocks the light from the WD photosphere, it makes 
%a shadow on the surface of the companion star.  Such an effect is also 
%included in our calculation.  
%The light curves are calculated for four cases of the companion mass, 
%i.e., $M_{\rm MS}= 0.8$, 1.1, 1.5 and $2.0 M_\odot$.
%Since we obtain similar light curves for all of these four cases,
%we show here only the results for $M_{\rm MS}= 1.5 M_\odot$.

\subsection{Accretion Disk Surface}
     We have included the luminosity coming from the accretion disk 
irradiated by the WD photosphere when the accretion disk
reappears several days after the optical maximum, 
i.e., when the WD photosphere shrinks to smaller than the size of
the inner critical Roche lobe.  Then we assume that the radius 
of the accretion disk is gradually increasing/decreasing to 
\begin{equation}
R_{\rm disk} = \alpha R_1^*,
\label{accretion-disk-size}
\end{equation}
in a few orbital periods,
where $\alpha$ is a numerical factor indicating the size of 
the accretion disk and $R_1^*$ is the effective radius of 
the inner critical Roche lobe for the primary WD component.
%given by Eggleton's (1983) formula.
The surface of the accretion disk absorbs photons 
and reemits the absorbed energy with a blackbody spectrum 
at a local temperature.  Here, we assume that 50\% 
of the absorbed energy is emitted from the surface 
while the other is carried into the interior of the accretion disk 
and eventually brought into the WD.
%,  mainly because a much smaller 
%efficiency such as 25\% results in a much larger disk radius
%to reproduce the brightness of the plateau phase.
The original temperature of
the disk surface is assumed to be constant at $T_{\rm ph, disk}= 4000$ K 
including the disk rim. 
The viscous heating is neglected because it is much smaller 
than that of the irradiation effects.
%This temperature is important on the side of the accretion disk
%because the side is never irradiated by the WD photosphere. 
\par
     We also assume that the accretion disk is axisymmetric and has 
a thickness given by 
\begin{equation}
h = \beta R_{\rm disk} \left({{\varpi} 
\over {R_{\rm disk}}} \right)^2,
\label{flaring-up-disk}
\end{equation}
where $h$ is the height of the surface from the equatorial plane,
$\varpi$ the distance on the equatorial plane 
from the center of the WD, 
and $\beta$ is a numerical factor showing the degree of thickness.
We have adopted a $\varpi$-squared law simply to mimic the effect
of the flaring up of the accretion disk rim 
(e.g., Schandl, Meyer-Hofmeister, \& Meyer 1997) %\cite{sch97})
and the exponent does not affect the disk luminosity so much 
mainly because the central part of the disk is not seen.
The surface of the accretion disk is divided into 32 pieces
logarithmically and evenly in the radial direction 
and into 64 pieces evenly in
the azimuthal angle as shown in Figure \ref{usco_fig80_w30}.
The outer rim of the accretion disk is also divided into 64 pieces
in the azimuthal direction and 8 pieces in the vertical direction
by rectangles.  
%When the photosphere of the WD becomes very small, e.g.,
%$R_{\rm disk}/R_{\rm ph} > 10$, we attribute the first 16 meshes 
%to the outer region (from $\varpi=R_{\rm disk}$ to 
%$\varpi=R_{\rm disk}/\sqrt{10}$) to avoid coarse meshes in the outer part, 
%and then 16 meshes to the inner region (from 
%$\varpi=R_{\rm disk}/\sqrt{10}$ to $\varpi=R_{\rm ph}$), 
%each region of which is divided logarithmically evenly.  
\par
\begin{figure}
\epsscale{.9}
%%%\plotone{fig3.eps}
\plotone{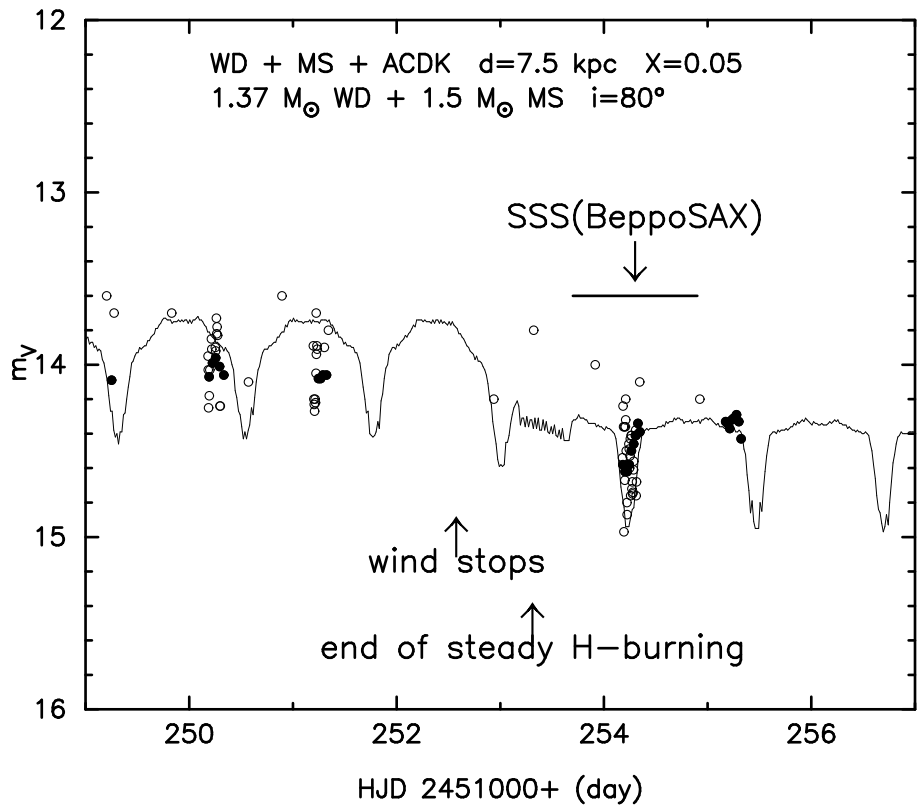}
%\plotfiddle{sn1a96let/evolution1.ps}{5.0cm}{270}{0.4}{0.4}{-170}{220}
\caption{
Same as Fig. 2, but for a part of the light curve that 
takes place during the transition 
from the wind phase to the cooling phase ($t \sim 14-22$ days).
When the optically thick wind stops at $t= 17.5$ days, 
the photosphere of the white dwarf envelope drastically shrinks 
from $\sim 0.1 R_\odot$ to $\sim 0.003 R_\odot$ within 1 day.  
This makes a drop $\Delta m_V \sim 0.3$ 
near $t \sim 18$ days (HJD 2,451,253), which is followed by 
a further drop $\Delta m_V \sim 0.4$ caused by a reduction 
of the accretion disk (ACDK) size.
The detection of an SSS by BeppoSAX is very consistent with 
the wind duration 
because supersoft X-rays emerge only after the massive wind stops. 
\label{vmag1370ecl_m15}}
\end{figure}

%     The luminosity of the accretion disk
%depends strongly on both the thickness $\beta$ and the size $\alpha$.
%We have examined 14 cases of $\alpha= 0.7$, 0.8, 0.9, 1.0, 1.1, 
%1.2, 1.3, 1.4, 1.5, 1.6, 1.7, 1.8, 1.9, and 2.0,
%and 10 cases of $\beta= 0.05$, 0.10, 0.15, 0.20, 
%0.25, 0.30, 0.35, 0.40, 0.45, and 0.50.
     We have reproduced the light curves in the quiescent phase 
($B$-magnitude, \cite{sch90}) by adopting
a set of parameters such as $\alpha=0.7$, $\beta=0.30$, 
a 50\% irradiation efficiency of the accretion disk, 
and $i=80 \arcdeg$ 
for the WD luminosity of $\sim 1000 L_\odot$
($A_B= 0.8$ and $d=15$ kpc, or $A_B= 2.8$ and $d=6$ kpc, see below),
which is roughly corresponding to the mass accretion rate 
of $\dot M_{\rm acc}= 2.5 \times 10^{-7} M_\odot$ yr$^{-1}$
for the $1.37 M_\odot$ WD.
%These $\alpha$ and $\beta$ parameters are 
%essentially similar to those for the supersoft X-ray source models 
%by Schandl et al. (1997).
\placefigure{vmag1370va1_m15}
\placefigure{vmag1370ecl_m15}

\section{RESULTS}
\begin{figure}
\epsscale{.9}
%%%\plotone{fig2.eps}
%%%\plotone{vmag1370va1_m15_c12d5.eps}
\plotone{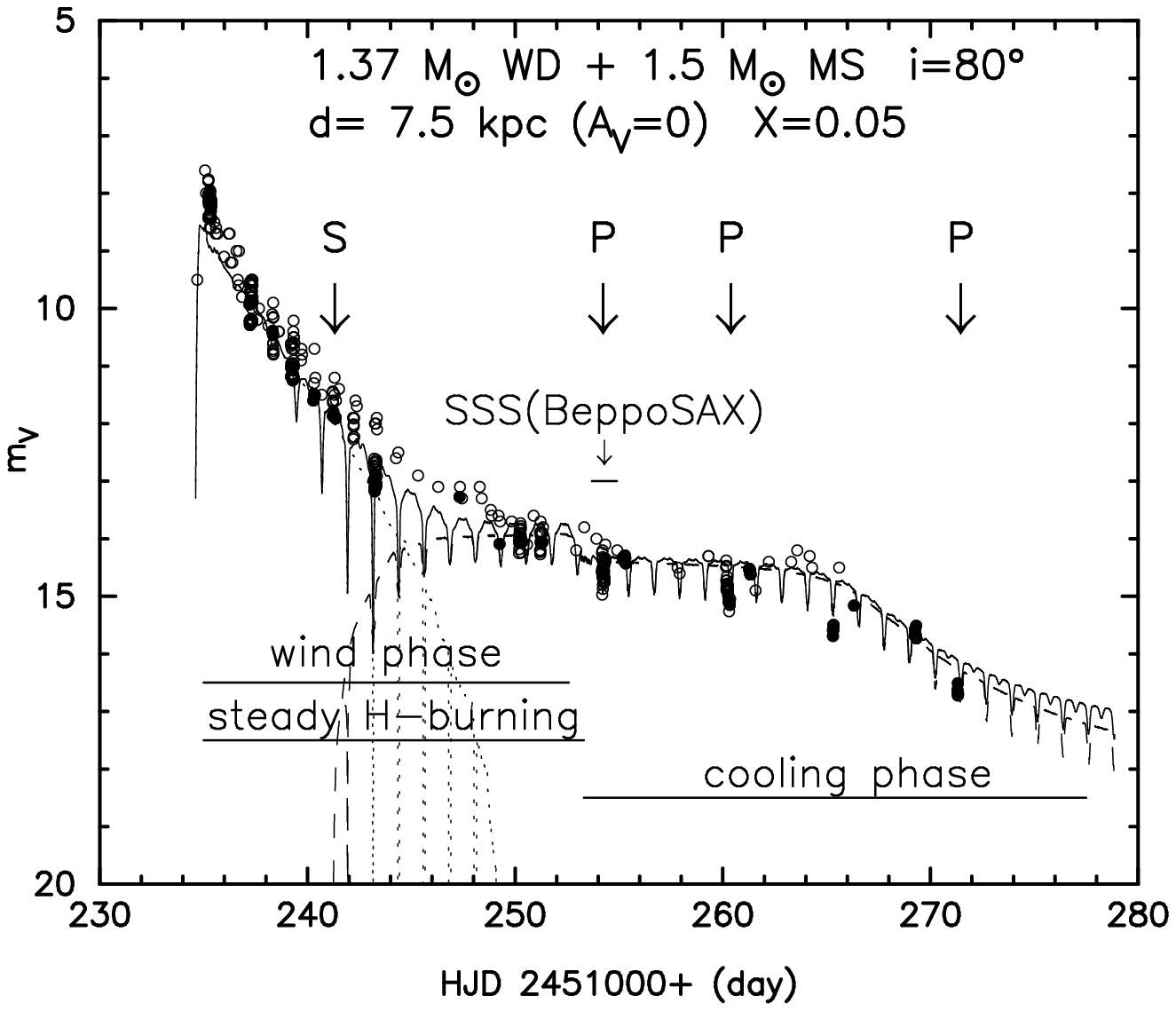}
%\plotfiddle{sn1a96let/evolution1.ps}{5.0cm}{270}{0.4}{0.4}{-170}{220}
\caption{
Theoretical light curve (solid line) plotted against time 
(HJD 2,451,000+), together with the observational points 
(the filled circles represent data taken from Matsumoto \& Kato 1999, 
the open circles from the VS-NET archives), 
for $1.37 M_\odot$ WD $+$ $1.5 M_\odot$ MS with
an inclination angle $i=80\arcdeg$. 
The brightnesses of the white dwarf photosphere (dotted line) and
of the accretion disk surface (dashed line) are also added.
The distance to U Sco is estimated to be 5.4---8.0 kpc for 
no absorption ($A_V=0$) but 4.1---6.1 kpc for $A_V=0.6$.  
U Sco was detected by BeppoSAX as a supersoft X-ray source (SSS) 
19---20 days after maximum (Kahabka et al. 1999).
The capital letters ``S'' and ``P'' mean the secondary minimum and
the primary minimum, respectively, in the eclipses observed 
by Matsumoto and Kato (1999).  
\label{vmag1370va1_m15}}
\end{figure}

     Figure \ref{vmag1370va1_m15} shows the observational 
points and
% (large open circles: taken from Matsumoto \& Kato 1999;
%small open circles; taken from the VS-NET summary).  
our calculated $V$-magnitude light curve 
(solid line). To maintain an accretion disk,
we assume a mass accretion rate 
$\dot M_{\rm acc}= 1 \times 10^{-7} M_\odot$ yr$^{-1}$ 
%in equation (\ref{dmdt_envelope_mass})
during the outburst.
\par
     To fit the early linear decay phase 
($t \sim 1-10$ days after maximum), we have calculated a total of 140
$V$-magnitude light curves for five cases of the WD mass:
$M_{\rm WD}= 1.377$, 1.37, 1.36, 1.35, and $1.3 ~M_\odot$, 
with various hydrogen content of $X=0.04$, 0.05, 0.06, 0.07, 0.08,
0.10, and 0.15 of the envelope, 
where the metallicity $Z=0.02$ is fixed, 
each for four companion mass of $M_{\rm MS}=0.8$, 1.1, 1.5, 
and $2.0 M_\odot$.   We choose $1.377 M_\odot$ 
as a limiting mass just before the SN Ia 
explosion in the W7 model ($M_{\rm Ia}=1.378 M_\odot$)
of Nomoto, Thielemann, \& Yokoi (1984).
\par
     We have found that the early 7 day light curve hardly
depends on the chemical composition and the companion mass but
is mainly determined by the white dwarf mass.
This is because (1) the early-phase light curve is determined mainly
by the WD photosphere (Fig. \ref{vmag1370va1_m15}, dotted line) 
and therefore by the wind mass loss rate and (2) the optically thick 
wind is driven by the strong peak of OPAL opacity, which is due 
not to helium or hydrogen lines but to iron lines (\cite{kat94}). 
Therefore, the determination of the WD mass is almost independent 
of the hydrogen content, the companion mass, or the disk configuration.
%It should be noted here that the accretion disk is not so important
%in this early 7 days because the accretion disk is engulfed 
%by the white dwarf envelope (dashed line in 
%Fig. \ref{vmag1370va1_m15}).  
The $1.37 M_\odot$ light curve is
in much better agreement with the observations
than are the other WD masses. 
%Therefore, we adopt the $M_{\rm WD}= 1.37~M_\odot$ in this Letter.
\par
     The distance to U Sco is estimated to be 5.4---8.0 kpc, as shown 
in Figure \ref{vmag1370va1_m15}, if we assume no absorption ($A_V=0$).
Here, we obtain 5.4 kpc for the fit to the upper bound and 
8.0 kpc for the fit to the lower bound of the observational points.
For an absorption of $A_V= 0.6$ (\cite{bar81}), we have the distance
of 4.1---6.1 kpc, and then U Sco is located 1.5---2.3 kpc 
above the Galactic plane ($b=22\arcdeg$).
These ranges of the distance are reasonable compared with 
the old estimates of the distance to U Sco such as 14 kpc 
(e.g., \cite{war95}).
\par
     To fit the cooling phase ($t \sim$ 30---40 days after maximum), 
we must adopt the hydrogen content of $X=0.05$ 
among $X=0.04$, 0.05, 0.06, 0.07, 0.08, 0.10, and 0.15 for 
$M_{\rm WD}= 1.37~M_\odot$.
This is because the hydrogen content $X$ is equivalent to
the mass of hydrogen burning in the envelope.
Therefore, $X$ determines the duration of hydrogen shell burning,
i.e., the duration of the midplateau phase.
%This is the reason why the hydrogen content $X$ is determined from
%the beginning of the cooling phase.
For $X=0.05$, the optically thick wind stops at $t= 17.5$ days, and  
the steady hydrogen shell-burning ends at $t= 18.2$ days.
This duration of the strong wind phase is very consistent with
the BeppoSAX supersoft X-ray detection 19---20 days after 
the optical peak (\cite{kah99}) because supersoft X-rays 
are self-absorbed by the wind itself during the strong wind phase.
It should be noted that we do not use this detection of the
supersoft X-rays to constrain any physical parameters.
% because of no observations before or after this detection. 
Hydrogen shell-burning begins to decay
from $t= 18.2$ days but still continues to supply a part 
of the luminosity; the rest comes from the thermal energy
of the hydrogen envelope and the hot ash (helium) below 
the hydrogen layer.  The thermal energy amounts to several times
10$^{43}$ ergs, which can supply a bolometric luminosity of 10$^{38}$ erg
s$^{-1}$ for ten days or so until $t \sim 30$ days, as seen in Figures 
\ref{vmag1370va1_m15} and \ref{vmag1370ecl_m15}.  
\par
     The envelope mass at the peak is estimated to be 
$\Delta M \sim 3 \times 10^{-6} M_\odot$ 
%in our solutions
for $M_{\rm WD}= 1.37 M_\odot$, $X=0.05$, and $Z=0.02$; 
thus, the average mass accretion rate of the WD 
%in the year 1987---1999 quiescent phase 
becomes $\sim 2.5 \times 10^{-7} M_\odot$ yr$^{-1}$ 
in the quiescent phase between 1987 and 1999, 
if no WD matter is dredged up into the envelope.
Such high mass accretion rates 
% as $\gtrsim 1 \times 10^{-7} M_\odot$ yr$^{-1}$ 
strongly indicate that the mass transfer is driven by
a thermally unstable mass transfer under the constraint 
that the companion star is a slightly evolved main-sequence star
(e.g., \cite{heu92}).   
%The thermally unstable mass transfer is realized only
%when the mass ratio is larger than 0.79, i.e.,
%$q= M_{\rm MS}/ M_{\rm WD} > 0.79$,
%which poses a requirement $M_{\rm MS} \gtrsim 1.1 M_\odot$.
\par
     The wind carries away about 60\% of the envelope mass, i.e.,
$\sim 1.8 \times 10^{-6} M_\odot$, which is much more massive 
than the observational indication of $\sim 1 \times 10^{-7} M_\odot$
in the 1979 outburst by Williams et al. (1981).  
The residual, $\sim 1.2 \times 10^{-6} M_\odot$, can accumulate 
in the WD. Therefore, the WD can grow in mass at an average rate of
$\sim 1.0 \times 10^{-7} M_\odot$ yr$^{-1}$.
%in the year 1987---1999.
%It should be noted that 
%the frictional mass ejection during the common envelope phase
%($t \sim 0-3$ days) is not important because the wind velocity 
%$\gtrsim 1000$ km s$^{-1}$ is much faster than 
%the orbital velocity $\lesssim 300$ km s$^{-1}$ 
%and the gas quickly leaves the binary without interacting with
%the orbital motion even inside the photosphere 
%(\cite{kat94}; \cite{kat99}). 
\par
     Our model fully reproduces the observational light curve 
if we choose $\alpha= 1.4$ and $\beta= 0.30$ during the wind phase
and $\alpha= 1.2$ and $\beta= 0.35$ during the cooling phase
for $1.37 M_\odot$ WD $+$ $1.5 ~M_\odot$ MS 
with $i\approx 80\arcdeg$.  
Since we obtain similar light curves for four companion masses, 
i.e., 0.8, 1.1, 1.5, and 2.0 $M_\odot$, 
we show here only the results of $M_{\rm MS}= 1.5 M_\odot$.
It is almost certain that
the luminosity exceeds the Eddington limit during the first day
of the outburst, because our model cannot reproduce 
the super Eddington phase.

\section{DISCUSSION}
     Eclipses in the 1999 outburst were observed by 
Matsumoto \& Kato (1999).  The depth of the eclipses is 
$\Delta V \sim 0.5-0.8$, and it is much shallower than that in quiescent 
phase ($\Delta B \sim 1.5$, \cite{sch90}).
Their observation also indicates almost no
sign of the reflection effect by the companion star.  
Thus, we are forced to choose a relatively large disk radius 
that exceeds the Roche lobe ($\alpha \sim 1.4$) 
and a large flaring-up edge of the disk ($\beta \sim 0.30$).
If we adopt a size of the accretion disk that 
is smaller than the Roche lobe, 
on the other hand, the companion star occults completely 
both the accretion disk and the white dwarf surface. 
As a result, we obtain a deep primary 
minimum as large as 1.5---2.0 or more.
%This is an inevitable result as long as we use the size of the 
%accretion disk smaller than the Roche lobe.
\par
\begin{figure}
\epsscale{.9}
%%%\plotone{fig1.eps}
\plotone{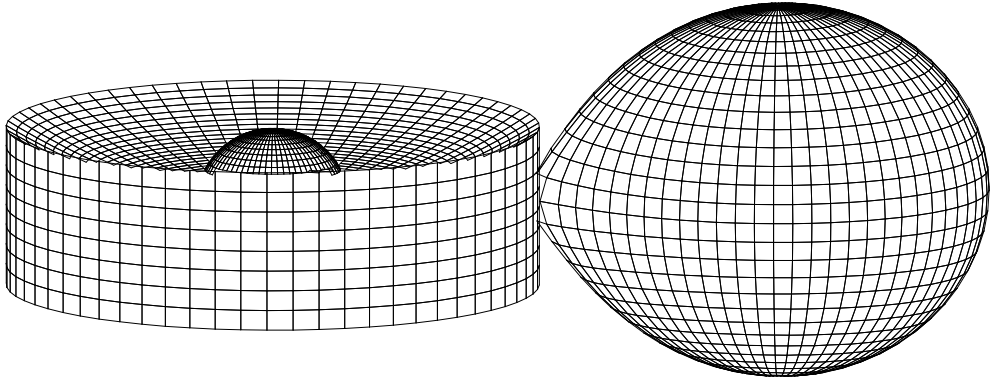}
%\plotfiddle{sn1a96let/dmdt_env1.ps}{5.0cm}{270}{0.4}{0.4}{-170}{220}
\caption{
Configuration of our U Sco 1999 outburst model $t \sim 9$ days 
after the maximum, i.e., when the WD photosphere shrinks to 
$R_{\rm ph} = 1.0 R_\odot$.
The cool component (right figure) is a slightly evolved MS
($1.5 M_\odot$) filling up its inner critical Roche lobe.  
Only the north and south polar areas of the secondary are 
heated up by the hot component ($1.37 ~M_\odot$ WD, left figure) 
because a large part of the light from the hot component 
is blocked by the flaring-up edge of the accretion disk.  
Here, the separation is $a= 6.87 R_\odot$, and
the effective radii of the inner critical Roche lobes are
$R_1^*= 2.55 R_\odot$, and $R_2^*= R_2= 2.66 R_\odot$, 
for the primary WD and the secondary MS, respectively.
%The photospheric radius of the hot component 
%is exaggerated in this figure to see it. 
\label{usco_fig80_w30}}
\end{figure}

     We have tested a total of 140 cases of the set ($\alpha, \beta$), 
i.e., 14 cases of $\alpha=0.7-2.0$ by $0.1$ step each for 10 cases of 
$\beta=0.05-0.50$ by $0.05$ step.  The best fit light curve 
is obtained for $\alpha=1.4$ and $\beta=0.30$.
Then a large part of the light from the WD photosphere is blocked 
by the disk edge, as shown in Figure \ref{usco_fig80_w30}. 
%When the WD photosphere shrinks much more than that in the figure.
%$R_{\rm ph} \sim 0.5 R_\odot$ at $t \sim 13$ days.
%Only the upper side of the accretion disk surface is seen 
%from Earth and this caldera-like visible area is lifted up 
%as the disk edge goes up.  So that the eclipse becomes shallower.
The large disk size and flaring-up edge may be driven 
by the Kelvin-Helmholtz 
instability because of the velocity difference between 
the wind and the disk surface.
After the optically thick wind stops, photon pressure may drive 
the surface flow on the accretion disk (\cite{fuk99}),
and we may expect an effect on the accretion disk similar to
what the wind has, but it may be much weaker.  
Thus, we may have a smaller radius of $\alpha=1.2$ but still have
$\beta=0.35$.  A much more detailed analysis of the results 
will be presented elsewhere.
\par
     It has been suggested that U Sco is a progenitor of SNe Ia 
(e.g., Starrfield, Sparks, \& Truran 1985)
because its WD mass is very close to the critical mass 
($M_{\rm Ia}= 1.38 M_\odot$) for the SN Ia explosion 
(\cite{nom84}).   
Hachisu et al. (1999) proposed a new evolutionary path 
to SNe Ia, in which they clarified the reason why 
the companion star becomes helium-rich even though 
it is only slightly evolved from the zero-age main sequence.
%(see also \cite{lih97}).   
%that such binary systems as U Sco are one of the 
A typical progenitor of SNe Ia in their WD+MS model, 
$M_{\rm 1,WD}=1.37 M_\odot$, $M_{\rm 2,MS} \sim 1.3 M_\odot$, and
$\dot M_{\rm 2,MS} \sim -2 \times 10^{-7} M_\odot$ yr$^{-1}$
just before the SN Ia explosion, has exactly the same system as U Sco.  
Thus, it is very likely that the WD in U Sco is now growing in mass
at an average rate of $\sim 1 \times 10^{-7} M_\odot$ yr$^{-1}$
toward the critical mass for the SN Ia explosion and will soon explode
as an SN Ia if the WD core consists of carbon and oxygen.
%Relatively high mass accretion rates 
%are maintained by the thermally unstable mass transfer 
%\par
%      Finally, it should be noted here that our main results 
%in this letter, (1) the WD mass is $\sim 1.37 M_\odot$ and very close 
%to the limiting mass for SN Ia explosion and (2) the WD is now growing
%in mass at the rate of $\sim 1 \times 10^{-7} M_\odot$ yr$^{-1}$,
%are derived from the early linear phase and from the beginning 
%of the cooling phase and, therefore, are almost independent of 
%the detailed assumptions of the accretion disk configurations.

\acknowledgments
     We are very grateful to many amateur astronomers who observed 
U Sco and sent their valuable data to VS-NET.  
We also thank the anonymous referee for many critical comments 
to improve the manuscript.
This research has been supported in part by the Grant-in-Aid for
Scientific Research (08640321, 09640325, 11640226, 20283591) 
of the Japanese Ministry of Education, Science, Culture, and Sports.
K. M. has been supported financially as a Research Fellow 
for Young Scientists by the Japan Society for the Promotion of Science.

\clearpage
%\epsfverbosetrue

% The \plotone and \plottwo commands scale the plot(s) in both dimensions
% so that the horizontal dimension fits in the body of the text.  The
% \plotfiddle command will override any automatic scaling, but often
% requires additional "fiddling" to get the plot to fit on the page.
% The \epsscale command allows the author to simply change the scaling
% of the plot in place, without the additional "fiddling" required by 
% \plotfiddle.

% That's all, folks.
%
% The technique of segregating major semantic components of the document
% within "environments" is a very good one, but you as an author have to
% come up with a way of making sure each \begin{whatzit} has a corresponding
% \end{whatzit}.  If you miss one, LaTeX will probably complain a great
% deal during the composition of the document.  Occasionally, you get away
% with it right up to the \end{document}, in which case, you will see
% "\begin{whatzit} ended by \end{document}".

\end{document}